\documentclass[letterpaper,twocolumn,english,aps,prl,superscriptaddress]{revtex4-1}
\usepackage{ae,aecompl}
\usepackage[T1]{fontenc}
\usepackage{textcomp}
\usepackage{amstext}
\usepackage{graphicx}
\usepackage[dvipdfm]{hyperref}
\hypersetup{
   colorlinks,
   linkcolor=blue,
   citecolor=blue,
   urlcolor=blue,
}
\usepackage[usenames, dvipsnames]{color}

\makeatletter

\newcommand{\lyxmathsym}[1]{\ifmmode\begingroup\def\b@ld{bold}
  \text{\ifx\math@version\b@ld\bfseries\fi#1}\endgroup\else#1\fi}

\@ifundefined{textcolor}{}
{%
 \definecolor{BLACK}{gray}{0}
 \definecolor{WHITE}{gray}{1}
 \definecolor{RED}{rgb}{1,0,0}
 \definecolor{GREEN}{rgb}{0,1,0}
 \definecolor{BLUE}{rgb}{0,0,1}
 \definecolor{CYAN}{cmyk}{1,0,0,0}
 \definecolor{MAGENTA}{cmyk}{0,1,0,0}
 \definecolor{YELLOW}{cmyk}{0,0,1,0}
 }

\makeatother

\usepackage{babel}
\begin{document}

\title{Pressure-tuning of bond-directional exchange interactions and magnetic frustration in hyperhoneycomb iridate $\beta$-$\mathrm{Li_2IrO_3}$}

\author{L. S. I. Veiga}

\affiliation{Deutsches Elektronen-Synchrotron (DESY), Hamburg 22607, Germany}

\affiliation{Advanced Photon Source, Argonne National Laboratory, Argonne, Illinois 60439, USA}

\affiliation{London Centre for Nanotechnology and Department of Physics and Astronomy, University College London, Gower Street, London, WC1E 6BT, United Kingdom}

\author{M. Etter}

\affiliation{Deutsches Elektronen-Synchrotron (DESY), Hamburg 22607, Germany}

\author{K. Glazyrin}

\affiliation{Deutsches Elektronen-Synchrotron (DESY), Hamburg 22607, Germany}

\author{F. Sun}

\affiliation{Advanced Photon Source, Argonne National Laboratory, Argonne, Illinois
60439, USA}

\affiliation{Center for High Pressure Science $\&$ Technology Advanced Research (HPSTAR), Shanghai, 201203, China }

\affiliation{Beijing National Laboratory for Condensed Matter Physics and Institute of Physics, Chinese Academy of Sciences, Beijing 100190, China}

\author{C. A. Escanhoela Jr.}

\affiliation{Advanced Photon Source, Argonne National Laboratory, Argonne, Illinois
60439, USA}

\affiliation{Brazilian Synchrotron Light Laboratory (LNLS), Brazilian Center for Research in Energy and Materials (CNPEM), Campinas, SP, 13083-970, Brazil}

\author{G. Fabbris}

\affiliation{Advanced Photon Source, Argonne National Laboratory, Argonne, Illinois
60439, USA}

\affiliation{Department of Physics, Washington University, St. Louis, Missouri
63130, USA}

\affiliation{Department of Condensed Matter Physics and Materials Science, Brookhaven National Laboratory, Upton, New York 11973, USA}

\author{J. R. L. Mardegan}

\affiliation{Deutsches Elektronen-Synchrotron (DESY), Hamburg 22607, Germany}

\author{P. S. Malavi}

\affiliation{Department of Physics, Washington University, St. Louis, Missouri
63130, USA}

\author{Y. Deng}

\affiliation{Department of Physics, Washington University, St. Louis, Missouri
63130, USA}

\author{P. P. Stavropoulos}

\affiliation{Department of Physics and Center for Quantum Materials, University of Toronto, 60 St. George St., Toronto, Ontario M5S 1A7, Canada}

\author{H.-Y. Kee}

\affiliation{Department of Physics and Center for Quantum Materials, University of Toronto, 60 St. George St., Toronto, Ontario M5S 1A7, Canada}
\affiliation{Canadian Institute for Advanced Research/Quantum Materials Program, Toronto, Ontario MSG 1Z8, Canada}

\author{W. G. Yang}

\affiliation{Center for High Pressure Science $\&$ Technology Advanced Research (HPSTAR), Shanghai, 201203, China }

\affiliation{High Pressure Synergetic Consortium (HPSynC), Geophysical Laboratory, Carnegie Institution of Washington, Argonne, Illinois 60439, USA}

\author{M. van Veenendaal}

\affiliation{Advanced Photon Source, Argonne National Laboratory, Argonne, Illinois
60439, USA}

\affiliation{Department of Physics, Northern Illinois University, De Kalb, Illinois
60115, USA}

\author{J. S. Schilling}

\affiliation{Department of Physics, Washington University, St. Louis, Missouri
63130, USA}

\author{T. Takayama}

\affiliation{Max Planck Institute for Solid State Research, Heisenbergstrasse 1, 70569 Stuttgart, Germany}

\affiliation{Department of Physics and Department of Advanced Materials, University of Tokyo, 7-3-1 Hongo, Tokyo, 113-0033, Japan }

\author{H. Takagi}

\affiliation{Max Planck Institute for Solid State Research, Heisenbergstrasse 1, 70569 Stuttgart, Germany}

\affiliation{Department of Physics and Department of Advanced Materials, University of Tokyo, 7-3-1 Hongo, Tokyo, 113-0033, Japan }

\author{D. Haskel}

\email{haskel@aps.anl.gov}

\affiliation{Advanced Photon Source, Argonne National Laboratory, Argonne, Illinois
60439, USA}

\date{\today{}}
\begin{abstract}
We explore the response of Ir $5d$ orbitals to pressure in $\beta$-$\mathrm{Li_2IrO_3}$, a hyperhoneycomb iridate in proximity to a Kitaev quantum spin liquid (QSL) ground state. X-ray absorption spectroscopy reveals a reconstruction of the electronic ground state below 2 GPa, the same pressure range where x-ray magnetic circular dichroism shows an apparent collapse of magnetic order. The electronic reconstruction, which manifests a reduction in the effective spin-orbit (SO) interaction in $5d$ orbitals, pushes $\beta$-$\mathrm{Li_2IrO_3}$ further away from the pure $J_{\rm eff}=1/2$ limit. Although lattice symmetry is preserved across the electronic transition, x-ray diffraction shows a highly anisotropic compression of the hyperhoneycomb lattice which affects the balance of bond-directional Ir-Ir exchange interactions driven by spin-orbit coupling at Ir sites. An enhancement of symmetric anisotropic exchange over Kitaev and Heisenberg exchange interactions seen in theoretical calculations that use {\em precisely} this anisotropic Ir-Ir bond compression provides one possible route to realization of a QSL state in this hyperhoneycomb iridate at high pressures.

\end{abstract}

\maketitle

The novel electronic ground states of $5d$-based compounds driven by spin-orbit interactions continue to provide an excellent playground for the realization of unconventional quantum phases of matter including topological insulators \cite{Hae15, Kim10, Kim12, Carter12} and quantum spin-liquids (QSLs) \cite{Khaliullin09, Balents14, Kimchi14}. One example of the latter is the non-trivial QSL ground state of the Kitaev model \cite{Kitaev06}, a rare example of a solvable interacting quantum model with Majorana fermions as its elementary excitations. Material candidates for possible realization of the Kitaev model include honeycomb-based-lattice systems with strong spin-orbit coupling \cite{Balents14, Rau15}, such as the two and three-dimensional honeycomb iridates, $\alpha$-$\mathrm{Li(Na)_2IrO_3}$ \cite{Chaloupka13, Cao13, Ye12, Singh12, Choi12, Hwan15, Knolle14}, $\beta$-$\mathrm{Li_2IrO_3}$ \cite{Takayama15, Biffin14, Schaffer15} and $\gamma$-$\mathrm{Li_2IrO_3}$ \cite{Modic14, Kimchi14, Glamazda16, Perreault15} as well as $\alpha$-$\mathrm{RuCl_3}$ \cite{Banerjee16, Winter16}. However, it is experimentally established that these materials order magnetically at low temperatures \cite{Takayama15, Modic14, Biffin14PRL, Biffin14, Williams16, Singh10}, spoiling numerous attempts to realize the Kitaev QSL. Hence, tuning structure and related intricate interactions present in these materials through chemical or physical pressure provides a potential route to introduce magnetic frustration and realize novel phases of matter.  

In this Letter we have investigated the electronic and structural response of $\beta$-$\mathrm{Li_2IrO_3}$ to high pressure. X-ray absorption near edge structure (XANES) measurements at Ir $L$-edges reveal a dramatic suppression of the isotropic Ir ($L_3/L_2$) branching ratio at $P \sim$ 1.5 GPa, signaling a reduction in the effective strength of spin-orbit interactions in the $5d$ band. This is the same pressure at which net magnetization in applied field collapses \cite{Takayama15}. The reconstructed electronic state preserves the $\langle L_z \rangle$/$\langle S_z \rangle$ orbital-to-spin moment ratio of Ir magnetic moments and the insulating ground state indicating that spin-orbit interactions and Mott physics continue to play a key role in driving the electronic ground state. The electronic/magnetic transition is driven by a highly anisotropic contraction of Ir-Ir bonds which alters the relative strength of direct and indirect hopping channels and related balance of bond-directional exchange interactions. Configuration interaction- and density functional theory- calculations corroborate that a strong interplay between hopping, Hubbard $U$ and spin-orbit effects is at play, facilitated by the rather large compressibility of this structure relative to that of other iridates (bulk modulus $B_0=100(8)$GPa). Remarkably, {\it ab initio} calculations on anisotropically compressed lattices based on $J$-$K$-$\Gamma$ spin hamiltonians ($J$-Heisenberg, $K$-Kitaev, $\Gamma$-symmetric anisotropic (SA) exchange interactions, respectively) \cite{Kim16} show that SA interactions become dominant at an effective pressure of $P\sim 1.4$ GPa. Since pure SA models lead to largely degenerate ground states in classical models \cite{Baek15, HY15} and quantum spin liquids in quantum models \cite{Rau14, Kim17}, the shift in the balance of bond-directional exchange interactions driven by anisotropic compression may explain the emergence of quantum paramagnetism and provides one possible route for realization of a novel QSL state in compressed $\beta$-$\mathrm{Li_2IrO_3}$.

\begin{figure}[t]
\centering
\includegraphics[width=1 \columnwidth]{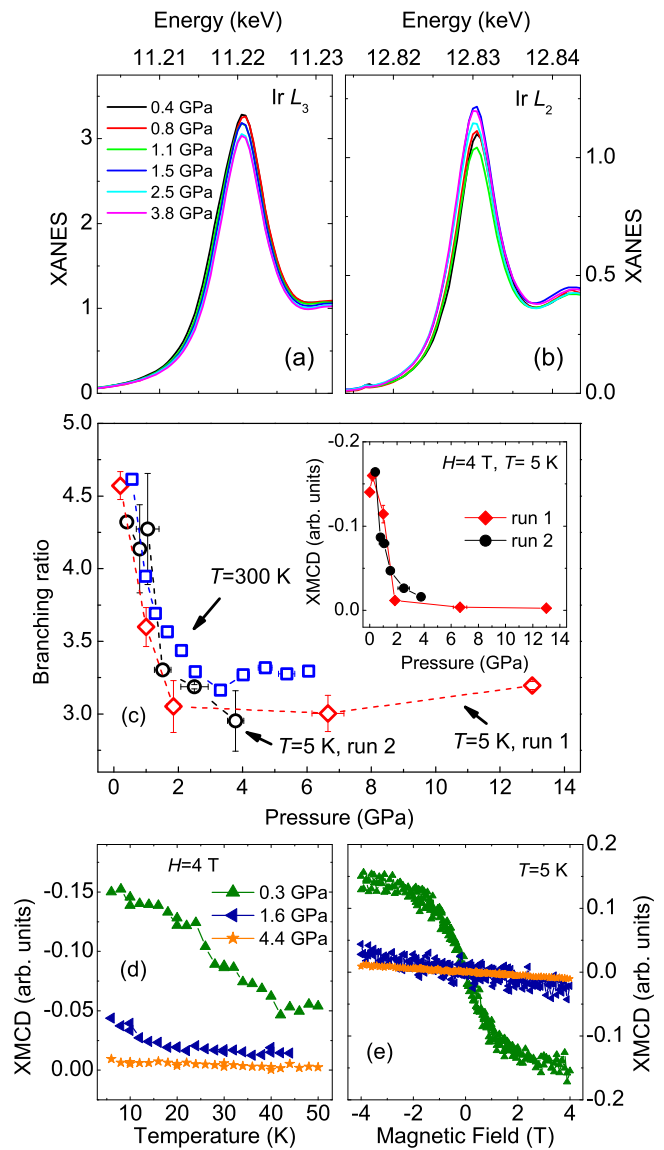}

\caption{(a,b) Ir $L_{2,3}$ XANES data at $T=5$ K as a function of pressure collected in experimental run 2. (c) Pressure dependence of the branching ratio at $T=5$ K and T=300 K measured in independent experiments. The inset shows the pressure dependence of the XMCD signal for two independent experimental runs (run 1 from Ref.\cite{Takayama15}). Note that the collapse of net magnetization coincides with the drop in BR. (d, e) Temperature- and field-dependent XMCD signal at selected pressures.} 
\label{fig:Fig1}

\end{figure}

The electronic and magnetic state of $\beta$-$\mathrm{Li_2IrO_3}$ was investigated through Ir $L$-edge XANES and x-ray magnetic circular dichroism (XMCD) measurements on polycrystalline samples at beamline 4-ID-D of the Advanced Photon Source of Argonne National Laboratory. Experimental details can be found in the Supplemental Material \footnote{See Supplemental Material at [URL will be inserted by publisher] for further information.}. Figure \hyperref[fig:Fig1]{1(a)-(b)} shows the isotropic x-ray absorption spectra at the iridium $L$ edges as a function of pressure. Of particular importance in the study of $4d$ and $5d$ oxides is the assessment of the relevance of spin-orbit interactions. The branching ratio, BR = $I_{L_3}/I_{L_2}$, is directly related to the ground-state expectation value of the angular part of the spin-orbit coupling, $\langle {\bf L} \cdot {\bf S} \rangle$, through BR = $(2+r) / (1-r)$, with $r= \langle {\bf L} \cdot {\bf S} \rangle / n_h$ and $n_h$ the number of holes in the $5d$ states \cite{Laan88}. Figure \hyperref[fig:Fig1]{1(c)} shows the pressure-dependence of BR obtained in three independent experimental runs. At ambient pressure, we measured BR = 4.5(1), which strongly deviates from the statistical value of 2, indicating the presence of a strong coupling between the local orbital and spin moments and proximity to a $J_{\rm eff}=1/2$ ground state \cite{Laguna-Marco10, Haskel12}. Under pressure, the BR decreases up to 2 GPa and maintains a constant value of $\sim3$ above 2 GPa. Using $n_h = 5$, $\langle {\bf L} \cdot {\bf S} \rangle$ changes from $2.27(2) \hbar^2$ at ambient pressure to $1.3(2)\hbar^2$ at 2 GPa. The reduction in BR coincides with the suppression of net magnetization in applied field as reported in Ref. \cite{Takayama15} and in the inset of Fig.~\hyperref[fig:Fig1]{1(c)}. Temperature- and field-dependent magnetization data, shown in Fig.~\hyperref[fig:Fig1]{1(d, e)} indicate possible emergence of quantum paramagnetism in the high pressure phase. Note that the drastic suppression of the BR accompanying the magnetic transition is distinct from what is observed for $\mathrm{Sr_2IrO_4}$ \cite{Haskel12} and $\mathrm{BaIrO_3}$ \cite{Laguna-Marco10, Laguna-Marco14}, where the BR remains intact through the collapse of the weak ferromagnetic ordering at $\sim17$ GPa and $\sim4.5$ GPa, respectively.

\begin{figure}[t]
\centering
\includegraphics[width=0.95 \columnwidth]{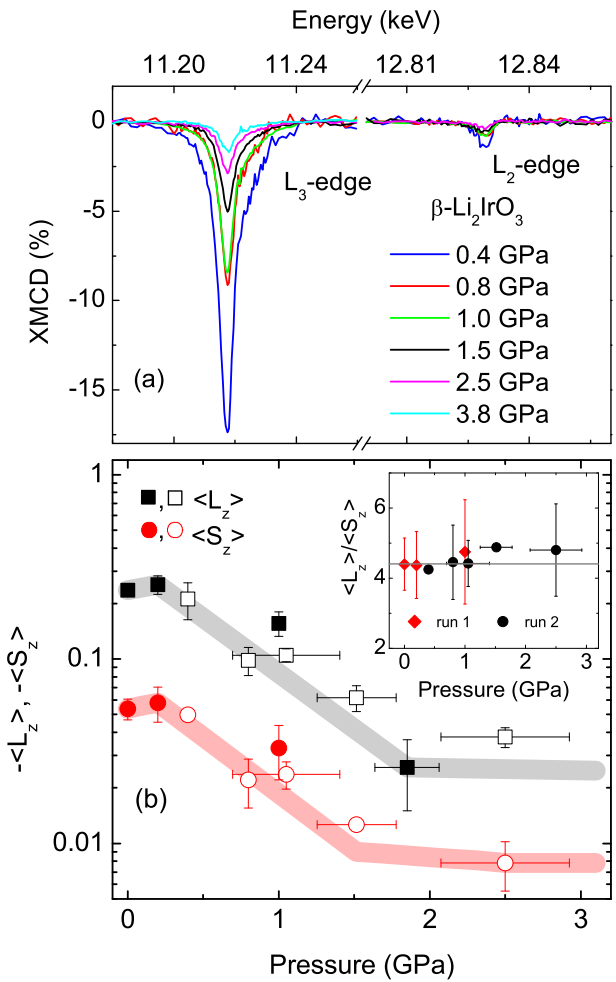}

\caption{(a) Normalized XMCD data at the Ir $L_{2,3}$-edges as a function of pressure for $\beta$-$\mathrm{Li_2IrO_3}$. The data were collected at $T=5$ K, $H=4$ T. (b) Pressure dependence of the ground-state expectation values of $L_z$ and $S_z$ for two independent experimental runs (run 1, closed symbols; run 2, open symbols). The inset shows the $\langle L_z \rangle / \langle S_z \rangle$ ratio as a function of pressure \cite{Note2, Takayama15}.}
\label{fig:Fig2}

\end{figure}

Additional information on spin-orbit coupling is provided by the ground-state expectation values of $L_z$ and $S_z$ extracted via sum rules analysis of the XMCD data at Ir $L_{2,3}$-edges \cite{Thole92, Carra93, Chen95}. The pressure-dependent XMCD data at both edges are shown in Fig.~\hyperref[fig:Fig2]{2(a)}. Noting that the number of holes in $5d$ states ($n_h=5$) is rather constant under pressure (sum of $L_{2,3}$ intensity in isotropic spectra does not vary more than 10\%), we have decomposed the Ir $5d$ moment into orbital and spin parts (see Fig.~\hyperref[fig:Fig2]{2(b)}) \footnote{The spin sum rule requires knowledge of the magnetic dipole operator {$\langle T_z \rangle$}. We have determined {$\langle T_z \rangle$} at ambient pressure using the spin sum rule with {$\langle S_z \rangle$} obtained by subtracting the orbital moment from the net magnetization (SQUID data) at 4 T. Since the point/symmetry space group does not change in the pressure range where the spin sum rule was applied, we used a pressure-independent {$\langle T_z \rangle / \langle S_z \rangle$} ratio in the spin sum rule to  calculate {$\langle S_z \rangle$} and {$\langle L_z \rangle / \langle S_z \rangle$} ratio as a function of pressure.}. Although the net orbital and spin magnetization is drastically suppressed by pressure, the orbital-to-spin moment ratio remains constant across the electronic/magnetic transition observed at $\sim 1.5$ GPa. The stability of $\langle L_z \rangle$/$\langle S_z \rangle$ indicates that the spin-orbit coupling in this material continues to play a key role in dictating the electronic ground state at high pressure.        


We now investigate the mechanism driving the reconstructed $5d$ state at high pressure. Since XANES probes all the empty $5d$ states, the measured $\langle {\bf L} \cdot {\bf S} \rangle$ includes contributions from a single hole in the $J_{\mathrm{eff}}=1/2$ state ($\langle {\bf L} \cdot {\bf S} \rangle \approx 1$) and 4 holes in the $e_g$ states ($\langle {\bf L} \cdot {\bf S} \rangle \approx 4 \times 3\zeta_{5d}/10Dq$). Here $\zeta_{5d}$ is the strength of the effective spin-orbit interactions and $10Dq$ the octahedral crystal field ($\zeta_{5d} \ll 10Dq$). Configuration interaction calculations indicate that a reduction in $\zeta_{5d}$ from 0.25 eV to 0.1 eV between pressures of 1.3 and 1.7 GPa can reproduce the BR data \cite{Note1} although a physical explanation for such strong reduction in $\zeta_{5d}$ in these atomic calculations is not apparent. Density functional theoretical calculations, which properly account for band effects, provide additional insight. As has been shown for both $\alpha$-RuCl$_3$ \cite{HY15} and $\beta$-$\mathrm{Li_2IrO_3}$ \cite{HSKim15}, electron correlations ($U_{\rm eff}=U-J_H$ where $U$ is on-site Coulomb repulsion and $J_H$ is Hund's coupling) have a significant impact on the effective strength of spin-orbit interactions in the $5d$ bands, hence on BR. Calculations on $\beta$-$\mathrm{Li_2IrO_3}$ with ambient pressure structure and $U_{\rm eff}$=2.5 eV (without magnetic order) give BR=4.32, while calculations using the $3.08$ GPa structure with $U_{\rm eff}$=1 eV give BR=3.45 (when considering magnetic order at ambient pressure and $U_{\rm eff}$=2.5 eV, BR changes to 4.66, see Supplemental Material for more details \cite{Note1}). Such a reduction in $U_{\rm eff}$ and concomitant reduction in the effective strength of SO interactions are driven by a change in Ir-Ir orbital overlap commensurate with the rather large compressibility of this structure, as discussed below. While a $J_{\rm eff}$ description remains valid, pressure pushes $\beta$-$\mathrm{Li_2IrO_3}$ away from the pure $J_{\rm eff}=1/2$ limit. A reduction in $\zeta_{5d}$ of $\sim$ 10\% can be obtained from the reduced separation between (predominant) $J_{\rm eff}=1/2, 3/2$ bands (also, $J_{\rm eff}=3/2$ character near the Fermi level increases from 16\% to 21\%, see \cite{Note1}). This is in good agreement with results in Ref.  \cite{HSKim15} where a 22\% reduction is seen when $U_{\rm eff}$ is reduced twice as much from 3.0 to 0.0 eV without a lattice contraction. Also, since $J_H \sim 0.5-1.0$ eV, a sizable $U\sim 1.5-2.0$ eV remains active in the high pressure phase explaining the preservation of the insulating gap as seen in transport measurements discussed below. The DFT calculations show a rather constant {$\langle L_z \rangle / \langle S_z \rangle$ $\approx 3.51$} in agreement with experiment. Since the lattice structure does not display discontinuities at the electronic transition, the suddenness of the BR collapse is likely a manifestation of the intricate interplay between $U$, $\zeta_{5d}$ and bandwidth that is a hallmark of this and other iridate systems. We note that other explanations for the BR drop, such as charge transfer from oxygen to Ir sites \footnote{In a hypothetical scenario, a transition to a fully occupied $J_{\mathrm{eff}}=1/2$ (Ir$^{3+}$, $5d^6$) state would also result in a sizable reduction in BR \cite{Laguna-Marco10}, but the required charge transfer energies are unphysical and we do not observe an energy shift of the leading absorption edge or peak absorption with pressure.}, strong deviations from octahedral symmetry, or Ir-Ir dimerization \footnote{Lattice symmetry is preserved across the electronic transition and lattice parameters do not show discontinuities. Lattice optimization calculations \cite{Kim16} show that dimerization of Ir-Ir Z-type bonds takes place only in the absence of SO and Coulomb interactions, which is not the case here. This was also observed for Ruthenate compounds \cite{HSKim16}. Furthermore, nearly constant {$\langle L_z \rangle / \langle S_z \rangle$} ratio and lack of measurable XANES energy shifts across the pressure-induced electronic/magnetic transition are inconsistent with a dimerization transition in agreement with XRD refinements.} can be ruled out by our data. We now turn to the structural response in order to seek further insight into the sudden electronic reconstruction and apparent collapse of magnetic order.


\begin{figure}[t]
\centering
\includegraphics[width=1. \columnwidth]{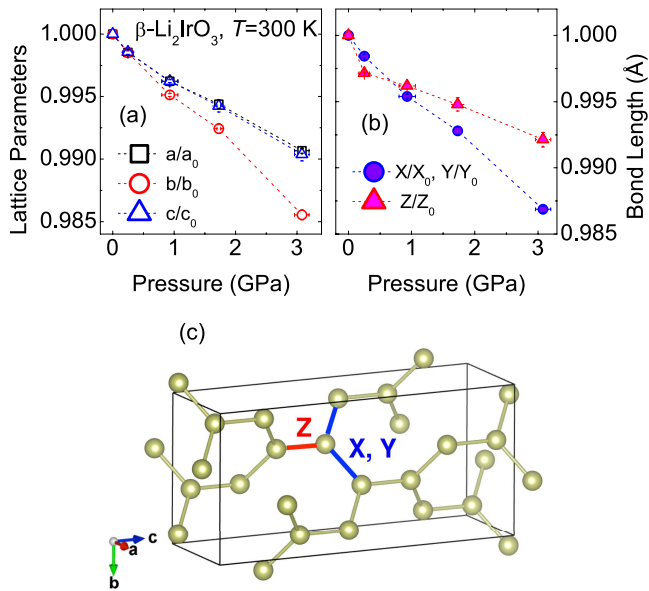}

\caption{Pressure-dependence of (a) lattice parameters and (b) X-, Y- and Z-bonds, all normalized to ambient pressure values. Note that in the $Fdddz$ space group, X and Y bond lengths are equal by symmetry. (c) Hyperhoneycomb structure of Ir atoms in $\beta$-$\mathrm{Li_2IrO_3}$. The blue X- and Y-bonds form the zigzag chains in the hyperhoneycomb network. The red bonds represent the Z-bonds which bridge the zigzag chains. The structure was visualized via VESTA software \cite{VESTA}.}
\label{fig:Fig3}
 
\end{figure}

Powder and single crystal x-ray diffraction (XRD) measurements were conducted at HPCAT beam line 16-BM-D of the Advanced Photon Source and P02.2 beam line of Petra III, respectively. Further details on the collection and analyses of the XRD data are given in the Supplemental Material \cite{Note1}. No structural phase transition is observed to 3.7 GPa which encompasses the electronic phase transition observed around 1.5 GPa. A new phase clearly emerges above 4.05 GPa \cite{Note1}. Lattice parameters were refined within the ambient pressure orthorhombic crystal structure (space group $Fdddz$) up to $P=3.7$ GPa. The pressure-dependent lattice parameters and Ir-Ir (X,Y,Z) bond lengths are shown in Fig. \hyperref[fig:Fig3]{3(a)-(b)}. The \emph{b} lattice parameter contracts at a faster rate than its \emph{a} and \emph{c} counterparts [$\frac{\Delta a/a_0}{\Delta P}=-0.30(1)$\%/GPa, $\frac{\Delta b/b_0}{\Delta P}=-0.47(1)$\%/GPa and $\frac{\Delta c/c_0}{\Delta P}=-0.31(2)$\%/GPa]. The faster \emph{b} axis compression leads to a nearly twofold increase in the compression rate of (X, Y) Ir-Ir bonds relative to Z-bonds (1.3\% vs. 0.7\% from 1 bar to 3 GPa; see Fig. \hyperref[fig:Fig3]{3(b)}). This is in striking agreement with theoretical calculations of optimized lattices in this material \cite{Kim16}, where (X,Y) and Z bonds contract by 3.4\% and 1.7\%, respectively at 10.2 GPa. The new crystal structure, persisting to the highest measured pressure of 8.5 GPa, was refined using single crystal data as having monoclinic symmetry, space group $C2/c$, and lattice parameters (at 5.8 GPa) $a=5.7930(8)$ \AA, $b=8.0824(16)$ \AA, $c=9.144(2)$ {\AA} and $\beta=106.777(15)^\circ$ \cite{Note1}. The first-order structural phase transition is accompanied by a volume collapse of $\sim$0.7\% at $\sim$4.4 GPa. The strain associated with the increasing anisotropy between $a$, $c$ and $b$ lattice parameters under pressure may trigger the transition to the high pressure, lower symmetry monoclinic phase \cite{Note1}.

One may be tempted to conclude that a shift of the structural transition to lower pressures on cooling drives the BR drop and collapse of net magnetization. However, the sudden reduction in BR occurs in the same pressure range at both $T=5$ K and $T=300$ K (Fig.~\hyperref[fig:Fig1]{1(c)}), almost $\sim 3$ GPa away from the onset of the structural phase transition. A small tetragonal distortion which gradually evolves within the low-pressure phase and changes sign across the structural phase transition does not affect the BR as seen experimentally and as verified by cluster calculations \cite{Note1}. In addition, $\beta$-$\mathrm{Li_2IrO_3}$ remains an insulator to at least 7 GPa, {\it i.e}., above both magnetic and structural transitions. While no signature of an insulator-metal transition is observed, the electronic gap (estimated from fits to the resistivity data) decreases linearly with pressure likely a result of a reduction in on-site Coulomb interactions upon pressure-induced increase in bandwidth \cite{Note1}. This is facilitated by a relatively low bulk modulus (100 GPa) relative to that of $\mathrm{Sr_3Ir_2O_7}$ (157 GPa) \cite{Donnerer16} and  $\mathrm{Sr_2IrO_4}$ (174 GPa) \cite{Haskel12}. If a collapse of local magnetic moment were to take place at the electronic/transition, one would expect a sudden change in Mott-Hubbard gap, which is not observed \cite{Takayama15}.
 

 


A recent high pressure study on the polytype $\gamma$-$\mathrm{Li_2IrO_3}$ reveals that the zero-field incommensurate spiral magnetic structure seen in both this and $\beta$ polytypes at ambient pressure is no longer present above $\sim1.5$ GPa \cite{Breznay17}, the same pressure where the (in-field) XMCD signal is strongly suppressed, pointing to a common ground state. Since XMCD probes net magnetization, it cannot directly rule out ordered phases with mute DC susceptibility. However, the strong suppression of the magnetic ordering temperature of $\beta$-$\mathrm{Li_2IrO_3}$ at 1.0 GPa (from $\sim38$ K to $\sim15$K) \cite{Takayama15} suggests vanishing of magnetic ordering at the electronic transition. This is confirmed by temperature- and field-dependent XMCD measurements which show no signs of magnetic ordering and a paramagnetic response in applied field (Fig. ~\hyperref[fig:Fig1]{1(d, e)}). These results point to a magnetically disordered state with strong magnetic correlations, {\it i. e.} a quantum paramagnet or QSL state. In the absence of strong magnetic correlations, one would observe a high magnetic susceptibility and sizable XMCD signal originating from a field-induced alignment of local moments. The XMCD signal of $\sim1.5\%$ observed at the $L_3$ edge in the magnetically disordered phase at $H=4$ T and $T=5$ K corresponds to a field-induced moment of about $\sim$0.04 $\mu_B$/Ir. This is inconsistent with an uncorrelated paramagnetic state which would display a nearly tenfold increase in ordered moment under such H/T conditions and provides strong support for the presence of interacting, localized moments which do not order as a result of frustrated exchange interactions. That the system remains insulating to 7 GPa lends further support to the presence of interacting local moments above 2 GPa.

The effect of pressure on lattice structure, local moment and intricate exchange couplings in $\beta$-$\mathrm{Li_2IrO_3}$ was recently investigated using \emph{ab initio} density functional theory calculations \cite{Kim16}. It is found that anisotropic compression of Ir-Ir bonds forming the hyperhoneycomb network significantly alters the relative strength of direct and indirect hopping channels between \emph{d} orbitals. In particular, a large increase in $t_{dd\sigma}$ hopping with pressure causes the bond-directional symmetric anisotropic (SA) exchange interaction ($\Gamma$ in $J$-$K$-$\Gamma$ spin hamiltonians \cite{Rau14}) to become dominant over Heisenberg ($J$) and Kitaev ($K$) interactions. Remarkably this crossover takes place at an (effective) pressure of $P=1.4$ GPa. It has been shown that pure SA interactions lead to a macroscopically degenerate manifold of classical ground states in hyperhoneycomb (3D) lattices \cite{Perkins17}, a signature of frustration \cite{Baek15}. Quantum calculations on finite size honeycomb lattices (2D) also confirm absence of magnetic order in pure SA models \cite{Perkins17}, and it was recently suggested that this ground state continuously connects to the Kitaev QSL in the presence of bond anisotropy \cite{Kim17}. Our single crystal XRD experiments at high pressure show the same type of (X,Y)- and Z-bond anisotropy seen in the theoretical calculations despite the lack of exact agreement in the compressibility of the lattice parameters. This lends support to the SA interaction model put forward in Ref.~\cite{Kim16} as one possible mechanism explaining the apparent disappearance of magnetic order in $\beta$-$\mathrm{Li_2IrO_3}$. Since the electronic reconstruction accompanies disappearance of magnetic order while keeping a finite charge gap (with no change in lattice symmetry), it is possible that a 3D spin liquid state is stabilized under increasing strength of bond-dependent SA interactions under pressure. Probing the electronic and magnetic ground state that emerges in the high-pressure phase with other techniques, such as resonant inelastic x-ray scattering or inelastic neutron scattering, may shed additional details on whether the magnetic excitations are indeed non-trivial as expected for a QSL state.


\begin{acknowledgments}
We thank A. Pakhomova for the single crystal experiment at beam line P02.2 of Petra III. We also thank H.-P. Liermann, A. Ehnes and I. Schwark for the use of ECSI/Petra III facilities. Work at Argonne is supported by the US Department of Energy, Office of Science, Office of Basic Energy Sciences, under Contract No. DE-AC- 02-06CH11357. L. S. I. V. was partly supported by FAPESP (SP-Brazil) under Contract No. 2013/14338-3 during the experiments at APS. Work at UCL is supported by the Engineering and Physical Sciences Research Council (EPSRC).  G. F. was supported by the U.S. Department of Energy, Office of Science, Office of Basic Energy Sciences, under Contract No. DE-SC00112704, and Early Career Award Program under Award No. 1047478. M. v. V. is supported by the U. S. Department of Energy (DOE), Office of Basic Energy Sciences, Division of Materials Sciences and Engineering under Award No. DE-FG02-03ER46097 and NIU Institute for Nanoscience, Engineering, and Technology. Research at Washington University was supported by the National Science Foundation (NSF) through Grant No. DMR-1104742 \& 1505345. P.P.S and H.-Y. K are supported by Natural Sciences and Engineering Research Council of Canada. Computations were mainly performed on the GPC supercomputer at the SciNet HPC Consortium. SciNet is funded by the Canada Foundation for Innovation under the auspices of Compute Canada, the Government of Ontario, the Ontario Research Fund-Research Excellence, and the University of Toronto. We also thank GSECARS for use of laser drilling facilities. Part of this research was carried out at PETRA III at DESY, a member of Helmholtz Association (HGF). 
\end{acknowledgments}

\bibliography{references}

\end{document}